\documentclass[11pt,a4paper]{article}

\usepackage{amsfonts}
\usepackage[dvips]{graphicx}
\usepackage{wrapfig}
\usepackage{amsthm}
\usepackage{amsmath}
\usepackage{epsfig}
\usepackage{t1enc}
\usepackage{amssymb}
\usepackage{latexsym}
\usepackage{bm}

\def\d{\delta}
\def\e{\epsilon}
\def\a{\alpha}
\def\b{\beta}
\def\c{\gamma}
\def\t{\theta}

\begin{document}

\renewcommand{\figurename}{\small Fig.}


\centerline{\LARGE \bf LIPS-thermalization of a relativistic gas}

\vskip 5mm \centerline{\large Vladim\' ir Balek\footnote{e-mail
address: balek@fmph.uniba.sk}}

\vskip 5mm \centerline{\large \it Department of Theoretical
Physics, Comenius University, Bratislava, Slovakia}

\vskip 1cm

{\small It is argued that two-particle collisions of relativistic
particles ''at a distance``, irrespective of their position in the
configuration space, generate uniform distribution of particles in
Lorentz invariant phase space.}

\vskip 5mm

{In a recent paper by Mere\v s at al. \cite{mer} a new numerical
procedure for generation of relativistic particles with the given
total energy-momentum is proposed. The procedure makes use of
two-particle collisions between particles considered in momentum
space only, with no definite positions in configuration space, and
seems to produce uniform distribution of particles in Lorentz
invariant phase space (LIPS) when applied repeatedly to a system
with non-uniformly distributed particles at the beginning. The
paper includes a sketch of an argument explaining why it is so. In
this note a detailed exposition of the argument is given.

In the procedure developed in \cite{mer}, collisions are linked
into sequences with two collisions per particle and the pairing of
particles in each sequence is governed by a single parameter whose
value is chosen at random. While this can be advantageous in
numerical calculations, it is technically simpler and physically
more instructive to consider a gas of particles colliding
completely at random. Our aim is to show that the equilibrium
distribution of particles in such a gas is uniform in LIPS, so
that the distribution necessarily becomes uniform {\it
asymptotically}, for the number of collisions approaching
infinity. If this is the case, the procedure with linked
collisions, which uses essentially the same mechanism, will yield
an asymptotically uniform distribution, too.

Consider a gas of $N$ nonidentical relativistic particles in which
there are perpetually going on elastic 2-particle collisions with
a given angular distribution of the outgoing particles in cms. The
particles are free between collisions and collide ''at a
distance``, irrespective of their actual position in the
configuration space. To describe the evolution of the gas we
divide it into steps with one collision per step, and pick the
pair of particles entering the collision in any given step at
random. Then we choose the directions in which the particles are
coming out of the collision in accord with their angular
distribution in cms. Both choices are obviously independent of the
choices made in previous steps, as well as of the sequence number
of the actual step. Thus, the evolution of the gas is a {\it
homogeneous Markov process}.

There is a vast body of literature on Markov processes, including
monographs devoted entirely to them (see, for example, \cite{str})
and books on probability theory where their properties are
discussed in detail. For a newcomer, the best choice is perhaps to
read an exposition on the subject in some lecture notes available
online, as are those cited in \cite{cer}.

A homogeneous Markov process is described by the {\it transfer
matrix} $P_{ij}$, whose element $(ij)$ is the probability that the
system passes in one step from the state $j$ to the state $i$.
Denote the probability that the system will be found in the $n$th
step in the state $i$ by $p_n (i)$. The probabilities in two
subsequent steps are related by the formula
$$p_{n + 1} (i) = \sum \limits_j P_{ij} p_n (j).$$
In statistical mechanics, one calls the states $i$, $j$, $\ldots$
''microstates``, and the states defined by the probability
distribution $p_n (i)$ ''ensemble states`` or ''macrostates``.
Equilibrium is a macrostate into which the system gets after
infinitely many steps from the initial macrostate with an
arbitrary probability distribution; in particular, an arbitrary
microstate (which can be regarded as a macrostate with one
probability equal to 1 and all others equal to 0). Thus, for the
probability distribution in equilibrium we have
\begin{equation}
p(i) = \sum \limits_j {\cal P}_{ij} p_0 (j), \quad {\cal P} = \lim
\limits_{n \to \infty} P^n. \label{eq:lim}
\end{equation}
On the other hand, equilibrium can be defined also as a macrostate
whose probability distribution stays constant during the evolution
of the system. Thus, the probability distribution in equilibrium
should satisfy
\begin{equation}
p(i) = \sum \limits_j P_{ij} p(j). \label{eq:eigen}
\end{equation}
In the first lecture notes cited in ref. \cite{cer}, the interplay
between equations (\ref{eq:lim}) and (\ref{eq:eigen}) is
demonstrated on a simple, but no less illuminating, example of
two-level system.

According to equation (\ref{eq:eigen}), the vector $p$ is an
eigenvector of the matrix $P$ with the eigenvalue 1. One can
easily see that there necessarily exists {\it some} vector with
this property, the reason being that the matrix $P$ obeys the
normalization condition
\begin{equation}
\sum \limits_i P_{ij} = 1. \label{eq:norm}
\end{equation}
Indeed, if we sum both sides of equation (\ref{eq:eigen}) over $i$
and use equation (\ref{eq:norm}), we arrive at identity; thus, the
determinant of the system is zero and the system has at least one
nontrivial solution (modulo rescaling). However, we would also
like to know that the solution is unique and nonnegative, and that
it can be expressed in the form (\ref{eq:lim}). It turns out that
all this is indeed the case provided the matrix $P$ is {\it
irreducible}; that is, either the matrix itself or some finite
power of it is positive (has all entries positive). In less formal
terms, the solution has desirable properties if we are able to
pass from any state to any other state in a finite number of
steps, so that we do not eventually become trapped in some
subspace of the state space. The proof is based on a well-known
theorem from the theory of nonnegative matrices, called {\it
Perron-Frobenius theorem}. More details can be found in
\cite{rob}.

The microstates of the gas are numbered by the $N$-tuples of
particle momenta $({\bf p}_1, {\bf p}_2, \ldots,$ ${\bf p}_N)$.
Thus, there are continuously infinitely many of them. However, we
can pass to the {\it discretized theory}, dividing the momentum
space into elementary cells of finite size. This is actually done
in numerical calculations. In the discretized theory, the number
of microstates is countable, and if we take into account that the
probability distribution we are interested in includes a
$\d$-function on total 4-momentum (which is replaced by
1/[4-dimensional volume of the order (size of the cell)$^4$]
$\times$ Kronecker delta after discretization), the number of
microstates becomes finite. As a result, previous considerations
can be applied without modifications to a gas of colliding
particles as a special case.

The basic building block of the theory is the expression for the
distribution of outgoing particles in a two-particle collision.
Let us compute this distribution in the laboratory frame in case
it is isotropic in cms. The quantities referring to cms will be
denoted by the index 0. The probability that the particle 1 comes
out of the collision in cms in the solid angle $d\Omega_{10}$ is
$$dP = \frac {d\Omega_{10}}{4\pi}.$$
Denote the magnitude of the momenta of ingoing particles in cms by
$\pi_0$, the momenta of outgoing particles in cms by $({\bf
p}_{10}, {\bf p}_{20})$, the total energy of ingoing particles in
cms by $\e_0$, and the energies of outgoing particles in cms by
$(E_{10}, E_{20})$. The energies are defined in terms of momenta
as $E_{10} = \sqrt{p_{10}^2 + m_1^2}$ and $E_{20} = \sqrt{p_{20}^2
+ m_2^2}$, where $m_1$ and $m_2$ are the masses of the colliding
particles. The expression for $dP$ can be rewritten as
$$dP = \frac 1{4\pi p_{10}^2} \d (p_{10} - \pi_0) d^3 p_{10} =
\frac 1{4\pi p_{10}^2} \d (p_{10} - \pi_0) \d ({\bf p}_{10} + {\bf
p}_{20}) d^3 p_{10} d^3 p_{20},$$ and if we use
$$\d (E_{10} + E_{20} - \e_0) = \frac 1{p_{10}(1/E_{10} + 1/E_{20})}
\d (p_{10} - \pi_0),$$ we find
$$dP = \frac {1/E_{10} + 1/E_{20}}{4\pi p_{10}} \d (E_{10} + E_{20}
- \e_0) \d ({\bf p}_{10} + {\bf p}_{20}) d^3 p_{10} d^3 p_{20}.$$
The product of $\d$-functions can be written as $\d^4 (p_{10} +
p_{20} - q_0)$, where $p_{10}^\mu = (E_{10}, {\bf p}_{10})$ and
$p_{20}^\mu = (E_{20}, {\bf p}_{20})$ are the 4-momenta of
outgoing particles in cms and $q_0^\mu = (\e_0, {\bf 0})$ is the
total 4-momentum of ingoing particles in cms. From the
conservation laws we also have $1/E_{10} + 1/E_{20} =
\e_0/(E_{10}E_{20})$ and $p_{10} = \pi_0$, hence
\begin{equation}
dP = \frac {\e_0}{4\pi \pi_0} \d^4 (p_{10} + p_{20} - q_0) \frac
{d^3 p_{10} d^3 p_{20}}{E_{10} E_{20}}. \label{eq:Pcms}
\end{equation}
The 4-dimensional $\d$-function as well as the element of LIPS by
which it is multiplied are relativistic invariants, so that they
can be both carried over into the laboratory frame simply by
wiping off the index 0. The factor in front of the $\d$-function
must be rewritten as an invariant, too, which means that we must
express $\pi_0$ in terms of $\e_0$ and subsequently replace $\e_0$
by $\sqrt{q^2}$ (invariant mass). By squaring the equation $\e_0 =
E_{10} + E_{20}$, separating the term $2E_{10} E_{20}$ and
squaring the resulting equation again we obtain
$$\e_0^4 - 2\e_0^2 (E_{10}^2 + E_{20}^2) + (E_{10}^2 - E_{20}^2)^2
= 0,$$ and if we insert here the expressions for $E_{10}$ and
$E_{20}$ with $p_{10} = p_{20} = \pi_0$ and solve for $\pi_0$, we
find
$$\frac {\pi_0}{\e_0} = \frac 12 \left(1 - \frac {2M^2}{\e_0^2} +
\frac {\mu^4}{\e_0^4} \right)^{1/2},$$ where $M^2 = m_1^2 + m_2^2$
and $\mu^2 = m_1^2 - m_2^2$. Inserting this in equation
(\ref{eq:Pcms}) and passing from cms to laboratory frame we
finally obtain
\begin{equation}
dP = \frac 1{2\pi} \left(1 - \frac {2M^2}{q^2} + \frac
{\mu^4}{q^4} \right)^{-1/2} \d^4 (p_1 + p_2 - q) \frac {d^3 p_1
d^3 p_2}{E_1 E_2}. \label{eq:P}
\end{equation}
The factor in front of the $\d$-function in the expression for
$dP$ serves as a normalization factor, making the total
probability equal to 1. The correct normalization of $dP$ is
guaranteed by the way the expression has been derived, and can be
easily verified in the nonrelativistic limit.

If the distribution of outgoing particles in cms is anisotropic,
we must insert a factor $f(\cos \theta_{10})$ into the expression
for $dP$, where $\theta_{10}$ is the angle between the initial and
final directions of motion of the particle 1 in cms and $f$ is a
nonnegative function satisfying
$$\int \limits_{-1}^1 f(x) dx = 2.$$ The argument of $f$
can be expressed in terms of $q^2$ and $q_1\ .\ p_1$, where $q_1$
and $p_1$ are the initial and final 4-momenta of the particle 1.
In particular, for $m_1 = m_2 = m$ we have
$$\cos \theta_{10} = \frac {q^2 - 4 q_1\ .\ p_1}{q^2 - 4m^2}.$$

Consider now a gas of $N$ particles with momenta $({\bf q}_1, {\bf
q}_2, \ldots, {\bf q}_N)$, and suppose that at the given moment
there occurs a two-particle collision, with all pairs of particles
colliding with equal probability. We are interested in the
probability of finding the gas after the collision in an
infinitesimal neighborhood of the point $({\bf p}_1, {\bf p}_2,
\ldots,$ ${\bf p}_N)$ of the $N$-particle momentum space. This can
be written as a sum of contributions of all pairs of particles,
\begin{equation}
dP = \frac 1{\cal N} \sum \limits_{a < b} dP_{ab} \prod \limits_{c
\ne a, b} \d ({\bf p}_c - {\bf q}_c) d^3 p_c, \label{eq:Ptot}
\end{equation}
where ${\cal N} = N(N - 1)/2$ is the number of pairs and $dP_{ab}$
is the probability that after the particles $a$ and $b$ have
collided, their momenta will be found in an infinitesimal
neighborhood of the point $({\bf p}_a, {\bf p}_b)$ of the
two-particle momentum space. Suppose for simplicity that the
distribution of outgoing particles in 2-particle collisions is
isotropic in cms. Then the probability $dP_{ab}$ is given by
equation (\ref{eq:P}) rewritten for particles $a$ and $b$,
\begin{equation}
dP_{ab} = \frac 1{2\pi} \left(1 - \frac {2M_{ab}^2}{q_{ab}^2} +
\frac {\mu_{ab}^4}{q_{ab}^4} \right)^{-1/2} \d^4 (p_{ab} - q_{ab})
\frac {d^3 p_a d^3 p_b}{E_a E_b}, \label{eq:Pab}
\end{equation}
where $p_{ab} = p_a + p_b$, $q_{ab} = q_a + q_b$, $M_{ab}^2 =
m_a^2 + m_b^2$ and $\mu_{ab}^2 = m_a^2 - m_b^2$. Taking into
account the $\d$-functions in equation (\ref{eq:Ptot}), we can
rewrite the $\d$-function in equation (\ref{eq:Pab}) as $\d^4 (p -
q)$, where $q^\mu$ and $p^\mu$ are the total 4-momenta of in- and
outgoing particles. We can also transform the noninvariant
3-dimensional $\d$-functions in (\ref{eq:Ptot}) into invariant
ones by multiplying them by the energies of the ingoing particles
$\e_c$. In this way we arrive at the formula
\begin{equation}
dP = F \d^4 (p - q) \prod \frac {d^3 p_c}{E_c}, \label{eq:Ptot1}
\end{equation}
with the distribution function $F$ defined as
\begin{equation}
F = \frac 1{\cal N} \sum \limits_{a < b} \frac 1{2\pi} \left(1 -
\frac {2M_{ab}^2}{q_{ab}^2} + \frac {\mu_{ab}^4}{q_{ab}^4}
\right)^{-1/2} \prod \limits_{c \ne a, b} \e_c \d ({\bf p}_c -
{\bf q}_c). \label{eq:distr}
\end{equation}
The probability distribution $dP$, regarded as a function of
initial and final momenta of the particles, is the transfer matrix
for our problem. It is continuous, but can be discretized by the
procedure mentioned earlier.

The gas in equilibrium is described by the probability
distribution
\begin{equation}
dp = f \d^4 (p - P) \prod \frac {d^3 p_c}{E_c}, \label{eq:p}
\end{equation}
where $p^\mu$ is the total 4-momentum of the gas, $P^\mu$ is the
value assigned to $p^\mu$ and $f$ is the distribution function in
equilibrium depending on the momenta $({\bf p}_1, {\bf p}_2,
\ldots, {\bf p}_N)$ of the particles forming the gas. Denote the
$N$-tuple of momenta $({\bf p}_1, {\bf p}_2, \ldots, {\bf p}_N)$
by $\hat p$. For the probability distribution $dp$ we have the
continuous version of equation (\ref{eq:eigen}),
$$dp(\hat p) = \int dP(\hat p, \hat q) dp(\hat q),$$
where we integrate over the $N$-tuple of momenta $\hat q$. By
inserting here from equations (\ref{eq:Ptot}) and (\ref{eq:p}) we
obtain
$$f(\hat p) \d^4 (p - P) = \int F(\hat p, \hat q) \d^4 (p - q)
f(\hat q) \d^4 (q - P) \prod \frac {d^3 q_c}{\e_c},$$ and if we
replace $q^\mu$ in the argument of the second $\d$-function on the
right hand side by $p^\mu$, using the fact that the value of
$q^\mu$ is fixed by the first $\d$-function, we find
\begin{equation}
f(\hat p) = \int F(\hat p, \hat q) f(\hat q) \d^4 (p - q) \prod
\frac {d^3 q_c}{\e_c}. \label{eq:eigen1}
\end{equation}

After having obtained an integral equation for the distribution
function $f$, our next goal is to prove that it is solved by
constant $f$. Thus, we want to show that $F$ satisfies the
equation
\begin{equation}
\int F(\hat p, \hat q) \d^4 (p - q) \prod \frac {d^3 q_c}{\e_c} =
1. \label{eq:id}
\end{equation}
Note that a similar equation with the integration over the
$N$-tuple of momenta $\hat p$,
$$\int F(\hat p, \hat q) \d^4 (p - q) \prod \frac {d^3 p_c}{E_c} =
1,$$ is just the normalization condition for $F$. In fact, it is a
continuous version of the normalization ``by columns`` introduced
in equation (\ref{eq:norm}). Our claim is that the transfer matrix
we have constructed is normalized also ''by rows``. (Strictly
speaking, the matrices in the two normalization conditions are not
the same, since they have different energies in the denominator.
[The elementary volumes referring to different momenta do not
matter, since they can be always chosen to have a fixed size
throughout the momentum space.] However, we can view both matrices
as different representations of a single matrix.)

To prove equation (\ref{eq:id}), we first transform it into an
equation for {\it two-particle} probability distribution. For that
purpose, we pick an arbitrary term $(a, b)$ in the sum over the
pairs of particles in (\ref{eq:distr}), rewrite the
energy-momentum conservation law so that only the 4-momenta of
particles $a$ and $b$ remain in it, and eliminate the
3-dimensional $\d$-functions by integrating over all $\bf q$'s
other than $({\bf q}_a, {\bf q}_b)$. In this way we obtain
\begin{equation}
\int \frac 1{2\pi} \left(1 - \frac {2M^2}{q_{12}^2} + \frac
{\mu^4}{q_{12}^4} \right)^{-1/2} \d^4 (p_{12} - q_{12}) \frac {d^3
q_1 d^3 q_2}{\e_1 \e_2} = 1. \label{eq:id1}
\end{equation}
This can be viewed as normalization of the two-particle transfer
matrix ``by columns''. To show that the equation is satisfied, it
suffices to replace $q_{12}^2$ in the expression in front of the
$\d$-function by $p_{12}^2$, rename $({\bf p}_1, {\bf p}_2)$ to
$({\bf q}_1, {\bf q}_2)$ and {\it vice versa}, and notice that the
resulting equation is just the normalization condition for the
probability $dP$ defined in equation (\ref{eq:P}). Thus, the
normalization of the two-particle transfer matrix ``by columns''
follows immediately from its normalization ``by rows''.

Previous reasoning generalizes trivially to anisotropic
collisions. In this case, the integral in equation (\ref{eq:id1})
contains an additional factor $f (\cos \theta_{10})$, but the
equation is still satisfied because the in- and outgoing momenta
enter the expression for $\theta_{10}$ symmetrically. This is seen
in the formula for $\cos \theta_{10}$ cited before, and can be
regarded as a consequence of the trivial fact that the angles by
which the vector ${\bf p}_{10}$ is deflected from the vector ${\bf
q}_{10}$ and {\it vice versa} are identical. Thus, equation
(\ref{eq:id}) is valid no matter what the distribution of outgoing
particles in cms.

We have shown that the condition of equilibrium is solved by the
probability distribution that is uniform in LIPS,
\begin{equation}
dp = C_P \d^4 (p - P) \prod \frac {d^3 p_c}{E_c}, \label{eq:p1}
\end{equation}
where $C_P$ is a normalization constant depending only on the
4-momentum of the gas $P^\mu$. Thus, the particles forming the
gas, if in equilibrium, {\it can be} distributed uniformly in
LIPS. However, we would like to be sure that they {\it are}
distributed uniformly in LIPS. In the discretized theory this is
guaranteed by the Perron-Frobenius theorem, provided the gas can
pass from any initial state to any final state after a finite
number of two-particle collisions. In fact, the final state does
not need to be varied, since the desired transition for any
initial and final state can be combined from one direct and one
inverse transition to some {\it reference} state. We will not
attempt to prove that the interpolating sequence of collisions
exists in general, but we will provide an illustrative example
showing that such sequence exists for a pair of states that are in
a sense maximally distant from each other, and that its length is
of order $N$.

Consider a many-particle gas with the total energy $E$ and zero
total momentum, and suppose for simplicity that the particles are
massless. We are interested in the transitions into the reference
state that is maximally randomized, with the energy of all
particles equal to $\e_r = E/N$ and the momenta filling uniformly
a sphere in the momentum space. If the gas is to evolve into this
state, the least favorable case seems to be if it is in the state
with a certain particle $A$ having energy ${\cal E}_0 = E/2$ and
momentum oriented in one direction, and all remaining particles
having energy $\e = {\cal E}_0/(N - 1) \doteq \e_r/2$ and momenta
oriented
\begin{figure}[ht]
\centerline{\includegraphics[height=3cm]{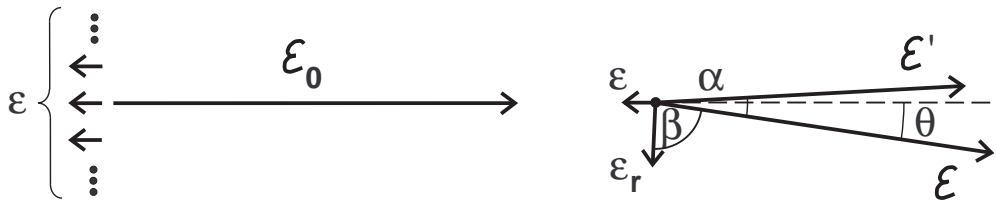}}
\caption{Randomization of a state with half the total energy
carried by a single particle} \label{fig:col}
\end{figure}
in the opposite direction (fig. \ref{fig:col} left). Consider a
sequence of $N - 1$ collisions in which the particle $A$ pairs
subsequently with all remaining particles, knocking them
alternately to one and the other side in a fixed plane and setting
their energy to $\e_r$. The energy-momentum conservation laws
written for the $\nu$th collision are (fig. \ref{fig:col} right)
\begin{equation}
{\cal E}' + \e_r = {\cal E} + \e, \quad {\cal E}' \cos \a + \e_r
\cos \b = {\cal E} - \e \cos \t, \quad {\cal E}' \sin \a + \e_r
\sin \b = \e \sin \t, \label{eq:con}
\end{equation}
where $\cal E$ and ${\cal E}'$ is the initial and final energy of
the particle $A$, $\a$ is the scattering angle of the particle
$A$, $\b$ is the angle between the initial momentum of the
particle $A$ and the final momentum of the particle $B$ that is
colliding with $A$, and $\t$ is the cumulative scattering angle of
the particle $A$ (the angle between the initial momentum of $A$ in
the $\nu$th collision and the momentum $A$ had before the first
collision).
In all scatterings but a couple of last ones the energy $\cal E$
is much greater than both $\e$ and $\e'$ (the final energy of the
particle $B$) $\doteq 2\e$. This causes $\a$ to be small and $\b$
to be close in absolute value to $\pi/2$. If we choose the first
particle $B$ to be declined from its original direction to the
right of the particle $A$, $\b$ will be of the form $\b = \mp \pi
+ \d$, where $\d$ is small and the upper and lower sign correspond
to odd and even collisions respectively. (In the figure, an odd
collision is depicted.) As for the third angle appearing in the
problem, for a sequence of collisions under consideration, with an
altering direction of motion of particles $B$ after collision,
$\t$ is always less in absolute value than the actual $\a$, hence
it is small, too.

We want to  solve equations (\ref{eq:con}) in the leading order in
the small parameter $\e/\cal E$. This is a simple exercise,
however, at one point we must be cautious: we can put $\e_r = 2\e$
almost everywhere, but in the term ${\cal E}' \cos \a$ in the
second equation (\ref{eq:con}), in which $\e_r$ appears through
${\cal E}'$, we must use the exact formula $\e_r = 2(1 - 1/N)\e$.
From the first equation (\ref{eq:con}) we obtain ${\cal E}' \doteq
\cal E - \e$, which yields
\begin{equation}
{\cal E}' \doteq {\cal E}_0 - \nu \e = \mu \e, \label{eq:en}
\end{equation}
where $\mu$ is the number of collisions which remain till the end.
For our purposes we can put $\mu = N - \nu$. (To obtain correct
values of ${\cal E}'$ at the last stage of the process, we must
pass to $\mu = N - 1 - \nu$ {\it and} account for the cumulative
effect of the small correction to $\e_r$.) From the remaining two
equations we find
\begin{equation}
\a = \pm \frac 2\mu, \quad \d = \pm \left( \frac 1\mu - \frac 1N
\right). \label{eq:ang}
\end{equation}
To compute $\t$, we must sum $\a$'s for all collisions up to that
with the sequence number $\nu - 1$. For odd collisions it holds
$$\t_{odd} = \frac 2{N - 1} - \frac 2{N - 2} + \ldots +  \frac 2\mu -
\frac 2{\mu + 1} = -2 \bigg[\frac 1{(N - 1)(N - 2)} + {}$$
$${} + \ldots + \frac 1{\mu (\mu + 1)} \bigg] \doteq - \int
\limits_\mu^{N - 1} \frac{dx}{x^2} = - \frac 1\mu + \frac 1{N - 1}
\doteq - \frac 1\mu + \frac 1N,$$ and for even collisions we have
$$\t_{even} = \underbrace{- \frac 1{\mu + 1} + \frac 1N}_{\t_{odd}(\nu - 1)} +
\frac 2{\mu + 1} = \frac 1{\mu + 1} + \frac 1N \doteq \frac 1\mu +
\frac 1N.$$ The momentum acquired by the particle $B$ after it
collided with the particle $A$ is deflected from the direction
perpendicular to the original momentum of the particle $A$ by the
angle $\Delta = \d + \t$. After inserting here for $\d$ and $\t$
we obtain
\begin{equation}
\Delta_{odd} = 0, \quad \Delta_{even} = \frac 2N.\label{eq:del}
\end{equation}

The approximate theory developed here holds for a majority of the
particles of the gas. After the collisions with the particle $A$
are completed, these particles split into two equally large
groups, one containing particles knocked to one side of the
particle $A$ and the other containing particles knocked to the
other side. From equation (\ref{eq:del}) we can see that the
particles of the first group move perpendicularly to the original
direction of motion of the particle $A$, while the particles of
the second group are slightly deflected to that side to which the
particle $A$ was heading at the beginning. A small third group
consists of particles that collided last and move in various
directions in such a way that their total momentum balances the
momentum arising from the deflection of the particles of the
second group.

Collisions with the particle $A$ represent just the first step of
a three-step procedure that brings the gas to the reference state.
The next step is to modify the direction of motion of the
particles so that they combine into pairs with exactly opposite
momenta; or in the discretized theory, with momenta that lie in
exactly opposite cells of the momentum space. If the size of the
cells $\Delta p$ is greater than $\Delta p_0 = \Delta_{even} \e_r
\doteq 4\e/N$, the task is restricted to the particles of the
third group and can be presumably completed in a number of steps
of order of the number of particles belonging to that group, which
is negligible in comparison to $N$. In the opposite case we must
take care of the first two groups of particles, too. We must
slightly rotate, say, the direction of motion of the particles of
the second group so that it will become opposite to the direction
of motion of the particles of the first group. This is easily
accomplished with one collision per particle; we just have to
think of it in advance and pick a slightly different final state
in the collisions of this group of particles with the particle
$A$. A straightforward computation yields that the final energy
$\e'$ of the particles of the second group must differ from the
energy $\e_r$ by the quantity
\begin{equation}
\Delta \e' = \frac {4\e\cos \phi}{N(1 + \sin  \phi + \cos \phi)},
\label{eq:dep}
\end{equation}
where $\phi$ is the angle by which the momentum of the particle of
the third group, paired with the given particle of the second
group, is deflected from the original momentum of the particle
$A$. Note that the expression for $\Delta \e'$ diverges for $\phi
= \pi$, therefore the procedure does not work with particles that
have not collided with the particle $A$ yet. To find how many
collisions are necessary, notice that we must straighten the
momenta of half the number of particles, and need one collision
for one straightening (perhaps a little more for the particles of
the third group, but their number is negligible anyway). Thus, the
number of collisions needed at this stage is $N/2$.

The third and final step of the randomization procedure is to
distribute the particles uniformly on a sphere in the momentum
space with the radius $\e_r$. This is done simply by colliding
particles with opposite momenta. The number of collisions is again
$N/2$, which makes the total number, in the less favorable case
when the cells in the momentum space have a size less than $\Delta
p_0$, equal to $2N$.

An isolated gas whose particles collide with each other wanders
the energy hypersurface of the $N$-particle momentum space. If the
hypersurface consists of domains that are interconnected but not
connected with each other, in the sense that points of one domain
cannot be reached from another domain, there can exist a
macrostate that stays unchanged once it was established, but the
gas will not evolve towards it from other macrostates. The
question is whether the uniform distribution in LIPS defined in
equation (\ref{eq:p1}) does not represent such a ``virtual''
equilibrium state. The example with randomization of the gas with
one highly energetic particle disputes this possibility since it
shows that even faraway points of the momentum space can be
accessible to each other.

For a given homogeneous Markov process, one can calculate the
equilibrium probability distribution numerically by starting with
an arbitrary state $i_0$, booking the state after every $\mu \gg
1$ steps and computing the relative frequency of every state after
$\nu \gg \mu$ steps. The probability distribution $p_{app} (i)$
obtained in this way is arguably a good approximation to $p(i)$.
To see that, consider the first sequence of $\mu$ steps. The
probability distribution $p_\mu (i)$ at the end of the sequence is
given by a cut-off version of equation (\ref{eq:lim}) with $p_0
(i) = \d_{ii_0}$,
$$p_\mu (i) = \sum \limits_j {\cal P}^{(\mu)}_{ij} p_0 (j) =
{\cal P}^{(\mu)}_{ii_0}, \quad {\cal P}^{(\mu)} = P^\mu.$$ For
large enough $\mu$ the information about the initial state is
effectively wiped out during the sequence, so that the effect of
the cut-off is small and $p_\mu (i)$ is close to $p(i)$. Thus, we
could find $ p_{app} (i)$ simply by calculating $p_\mu (i)$
numerically; that is, by repeating the sequence with the same
initial state sufficiently many times and computing the relative
frequencies of the final states. However, if we picked by chance
an initial state for which the system would wander through
marginal regions of the state space for a long time (this can
presumably happen for initial states with small probability), the
resulting probability distribution would be distorted. To be on
the safe side, we can use the procedure outlined at the beginning,
with the final state of one sequence serving as the initial state
of another sequence. This can be described by the formula
$$p_{app} (i) = \sum \limits_j {\cal P}^{(\mu)}_{ij} p_{app} (j).$$
The same formula is valid, as a trivial consequence of equation
(\ref{eq:eigen}), for $p(i)$. Nevertheless, $p_{app} (i)$ is still
an approximation, since the number of events used in its
computation is finite. For a relativistic gas, numerical
calculations following this recipe support our claim that the
equilibrium distribution of particles is uniform in LIPS
\cite{cer1}.

To complete the discussion, consider a gas with linked collisions
introduced in \cite{mer}. Suppose the two-particle collisions come
in sequences consisting of $N$ collisions, with the sequence
number of colliding particles, regarded as a periodic variable
with the period $N$, raised by one in each consecutive collision.
Furthermore, suppose that the difference between the sequence
numbers $d$, which stays constant throughout each sequence,
assumes all values 1, 2, $\ldots$, $N - 1$ with the same
probability. The collisions fall into three classes (fig.
\ref{fig:str}): collisions No. 1 to $D = $ min$(d, N - d)$ in
which both incoming particles are in the initial state (class
$\a$), collisions No. $D + 1$ to $N - D$ (absent for $N$ even and
$D = N/2$)
\begin{wrapfigure}[14]{l}{7.4cm}
\includegraphics [width=7.2cm]{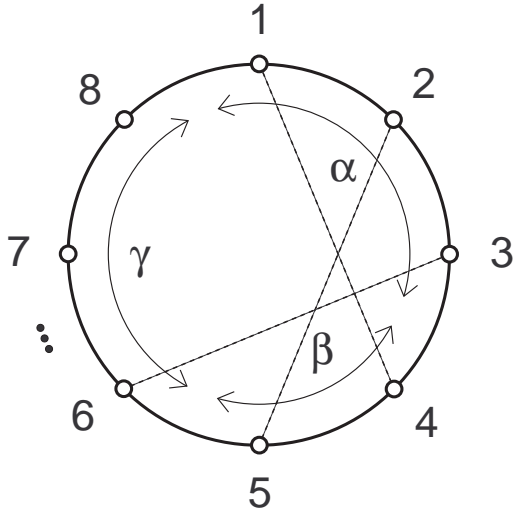}
\caption{Pairing of particles in collisions} \label{fig:str}
\end{wrapfigure}
in which one incoming particle is in the initial state and the
other incoming particle is in the intermediate state (class $\b$),
and collisions No. $N - D + 1$ to $N$ in which both incoming
particle are in the intermediate state (class $\c$). The
two-particle probability distributions in the respective classes
are $dP^{(\a)}= dP(qq,ll)$, $dP^{(\b)}= dP(ql,lp)$ and $dP^{(\c)}=
dP(ll,pp)$, where $q$, $l$ and $p$ are initial, intermediate and
final momenta respectively. The $N$-particle probability
distribution for one sequence of collisions is
\begin{equation}
dP = \frac 1{N - 1} \sum dP_d, \label{eq:ps}
\end{equation}
where the contribution of collisions with the given shift in the
sequence number can be written as a product of $N$ two-particle
probability distributions integrated over the intermediate
momenta,
\begin{equation}
dP_d = \int dP^{(\a)}_{1, D + 1} \ldots dP^{(\b)}_{D + 1, 2D + 1}
\ldots dP^{(\c)}_{N - D + 1, 1} \ldots \label{eq:psd}
\end{equation}
The $N$-particle distribution is again of the form
(\ref{eq:Ptot1}), and is again normalized ``by columns'' thanks to
the fact that the two-particle distribution is normalized ``by
columns''. (Equation (\ref{eq:id}) for the distribution function
reduces again to the identity (\ref{eq:id1}).) Thus, the linking
of collisions does not affect our conclusion that the uniform
distribution in LIPS obeys the condition of equilibrium. The
integrations in (\ref{eq:psd}) remove the better part of the
$\d$-functions in the two-particle probability distributions,
leaving us with a much greater freedom in choosing the final
momenta than we had in the case with independent collisions.
Because of that it could be expected that in a many-particle gas
the particles will need only $O(1)$ sequences of collisions to get
from any state to any other state. This, if true, would make the
total number of intermediate collisions of order $N$, just as we
have established in the example with randomization. However, to
prove that seems to be no less difficult for linked collisions
than for independent ones. On the other hand, if the particles
could indeed pass from any state to any other state within $O(N)$
independent collisions, they could of course do that also within
$O(N)$ {\it sequences} of collisions. As a result, if the true
(not ``virtual'') equilibrium distribution is uniform in a gas
with independent collisions, it is necessarily uniform also in a
gas with linked collisions.

\vskip 2mm {\it Acknowledgement.} I am grateful to Vladim\'ir \v
Cern\'y for stimulating discussions.

}

\enddocument